\documentclass[sigconf,authordraft,nonacm,review=false,timestamp=false]{acmart}
\AtBeginDocument{%
  }

\setcopyright{acmlicensed}
\copyrightyear{2025}
\acmYear{2025}
\acmDOI{XXXXXXX.XXXXXXX}
\acmConference[Ubicomp '25]{Make sure to enter the correct
  conference title from your rights confirmation email}{October 12--16,
  2025}{Espoo}
\acmISBN{978-1-4503-XXXX-X/2018/06}

\newcommand{\freqcut}{4000\,Hz}
\newcommand{\freqcarrier}{30\,kHz}
\newcommand{\systemname}{UltrasonicSpheres}

\newcommand{\abs}[1]{|#1|}

\usepackage{todonotes}
\usepackage{balance}

\begin{document}

\title{\systemname: Localized, Multi-Channel Sound Spheres Using Off-the-Shelf Speakers and Earables}

\author{Michael K\"{u}ttner}
\email{michael.kuettner@kit.edu}
\orcid{0009-0000-9021-0359}
\affiliation{%
  \institution{Karlsruhe Institute of Technology}
  \city{Karlsruhe}
  \country{Germany}
}

\author{Valeria Zitz}
\email{valeria.zitz@kit.edu}
\orcid{0009-0004-1158-861X}
\affiliation{%
  \institution{Karlsruhe Institute of Technology}
  \city{Karlsruhe}
  \country{Germany}
}

\author{Kathrin Gerling}
\email{kathrin.gerling@kit.edu}
\orcid{0000-0002-8449-6124}
\affiliation{%
  \institution{Karlsruhe Institute of Technology}
  \city{Karlsruhe}
  \country{Germany}
}

\author{Michael Beigl}
\email{michael.beigl@kit.edu}
\orcid{0000-0001-5009-2327}
\affiliation{%
  \institution{Karlsruhe Institute of Technology}
  \city{Karlsruhe}
  \country{Germany}
}

\author{Tobias R\"{o}ddiger}
\email{tobias.roeddiger@kit.edu}
\orcid{0000-0002-4718-9280}
\affiliation{%
  \institution{Karlsruhe Institute of Technology}
  \city{Karlsruhe}
  \country{Germany}
}

\renewcommand{\shortauthors}{K\"{u}ttner et al.}

\begin{abstract}
  We present a demo of \systemname{}, a novel system for location-specific audio delivery using wearable earphones that decode ultrasonic signals into audible sound. Unlike conventional beamforming setups, \systemname{} relies on single ultrasonic speakers to broadcast localized audio with multiple channels, each encoded on a distinct ultrasonic carrier frequency. 
  Users wearing our acoustically transparent earphones can demodulate their selected stream, such as exhibit narrations in a chosen language, while remaining fully aware of ambient environmental sounds.
  The experience preserves spatial audio perception, giving the impression that the sound originates directly from the physical location of the source. This enables personalized, localized audio without requiring pairing, tracking, or additional infrastructure. Importantly, visitors not equipped with the earphones are unaffected, as the ultrasonic signals are inaudible to the human ear.
  Our demo invites participants to explore multiple co-located audio zones and experience how \systemname{} supports unobtrusive delivery of personalized sound in public spaces.
\end{abstract}


\begin{teaserfigure}
  \includegraphics[width=\textwidth,page=2,clip,trim=0cm 10.3cm 0cm 0cm]{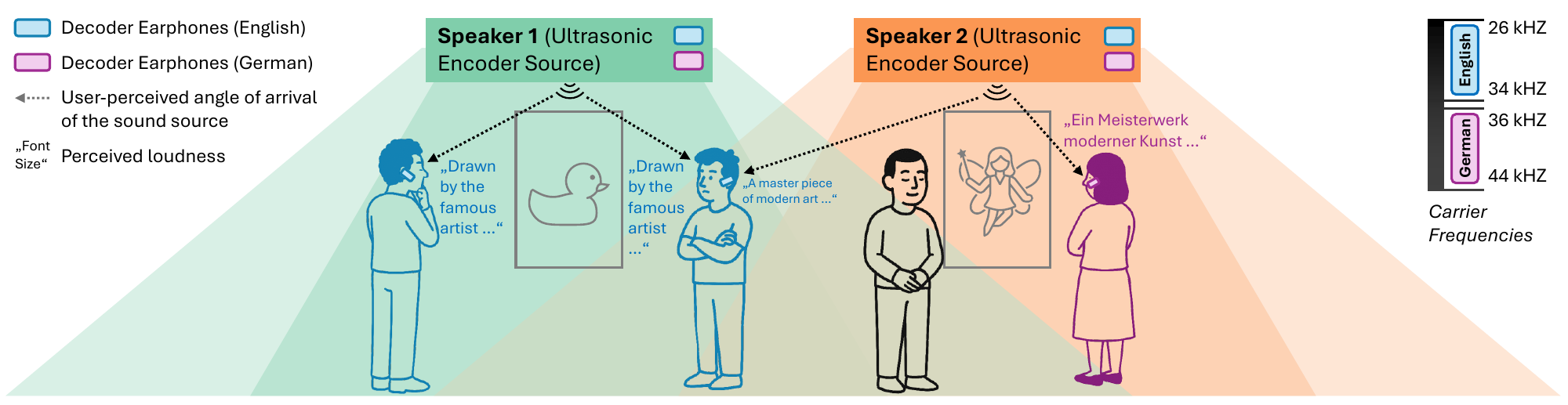}
  \caption{\systemname{} enables location-specific audio delivery using ultrasonic sources. The system transmits inaudible ultrasound that is converted by users' earphones into audible audio. This preserves spatial awareness and loudness perception, as if the sound was emitted audibly by the location-specific speaker. By encoding different audio channels (e.g., languages) on separate ultrasonic bands, \systemname{} allows multiple streams to be broadcast simultaneously from a single source. Users can then select their preferred stream directly on the earable device. Users without the corresponding earphones remain unaffected as they cannot hear the emitted ultrasound.}
  \label{fig:teaser}
\end{teaserfigure}

\received{30 May 2025}

\maketitle

\section{Introduction}

``Sound zones'' enable multiple co-located listeners to experience different audio content simultaneously without cross-interference \cite{betlehem2015personal}. These zones are especially useful in environments such as museums, libraries, and open offices, where personalized or context-aware audio enhances user experience without disturbing others. One promising approach to implementing sound zones involves the use of ultrasonic audio beams to project sound into precisely defined spatial regions.
Commercial solutions like the Holosonic Audio Spotlight use modulated ultrasound to produce highly directional, laser-like audio beams \cite{Holosonics}. However, such systems are expensive and proprietary. Research alternatives include phased arrays of ultrasonic transducers~\cite{ochiai2017holographic} and acoustic metasurfaces~\cite{zhong2025audible}, which offer precise control over sound localization. Despite their capabilities, these approaches require complex hardware, such as large transducer arrays or custom-fabricated components, limiting their accessibility and scalability.

In this demo paper, we introduce \systemname{}, a low-cost system that creates spatially confined ``sound spheres'' using commodity speakers and earphones. The core concept of \systemname{} is to transmit audio modulated onto inaudible ultrasonic carriers, which propagate through the air and are demodulated into audible sound by the earphone hardware. The system first modulates the desired audio content onto ultrasonic signals, which are emitted by strategically placed speakers. Compatible earphones can then demodulate these signals into clear, audible audio.
Since the signal is independently demodulated in each ear, the interaural time difference (ITD) is preserved, thus perceptually retaining the source's location.
Furthermore, \systemname{} support multiple concurrent channels by assigning each channel to a separate ultrasonic carrier frequency. For instance, this allows the delivery of different languages within the same physical zone.
Because ultrasonic signals exhibit strong spatial directivity, users can transition between different \systemname{} zones simply by moving through space. As they enter a sound sphere, their earphones demodulate the corresponding ultrasonic signal and play the associated audio; as they move away, the signal fades and another zone can take over. This enables an intuitive, interface-free method of audio selection based on physical location and orientation.

Attendees of the demo will be able to spatially experience different localized \systemname{} of two museum-like exhibits. They will experience how the system retains spatial perception and allows switching between different channels (e.g., switching between English and German narration).

\section{Related Work}
We first discuss localized spatial audio in public spaces. Then we summarize the use of ultrasound in earable research, as well as earphone-based sound spheres.

\subsection{Localized Spatial Audio in Public Spaces}
Spatial audio research has developed techniques that confine audible content to small regions. A common method is to use parametric ultrasonic arrays to produce ultrasound beams at high intensities and leveraging the air’s non-linearity to demodulate the carrier, producing audible sound only in the beam path \cite{mei2022parametric}. For example, Audio Spotlight \cite{Holosonics} is a commercial system that applies the approach to deliver localized audio in museum exhibits.
However, parametric speakers and their high-voltage amplifiers are specialized, costly and power-hungry. To add more flexibility, researchers have applied phased-array control to produce multiple highly localized sounds. For example, Ochiai et al.'s Holographic Whisper \cite{ochiai2017holographic} steers phase delays across an ultrasonic array to focus energy at arbitrary 3D positions. Each focal spot self-demodulates into a millimetre-scale virtual loudspeaker, and many such spots can be toggled independently to whisper different messages to people in the same space. Further, \citet{zhong2025audible} published ``audible enclave''. The system uses the intersection of two self-bending beams shaped by 3D-printed metasurface lenses. Neither beam is audible in flight, but where they cross, a nonlinear interaction generates sound, creating a pocket that only the listener can hear from within. While these systems demonstrate precise spatial targeting, they require custom hardware, complex calibration, and fixed installations.

\systemname{} addresses this limitation by requiring only a single ultrasonic source and demodulating earphones to deliver audio that is perceivable only within a confined spatial region. 

\subsection{Earables and Ultrasound}
In earable research (i.e., research that explores sensing in and around the ear), ultrasound has been used as a probing signal for different applications from ear canal based authentication \cite{gao2019earecho}, to mid-air gesture recognition \cite{jin2021sonicasl}, and heart rate tracking \cite{fan2023apg}. Most closely related to our work, Watanabe et al. \cite{watanabe2023ultrasonicwhisper} have shown that ultrasonic transmitters can be used for malicious interference with earphones by injecting commands into a victim’s noise canceling earbud, which demodulates them into audible messages. While this poses a potential security threat, it also shows the ability of commodity ear-worn devices to receive ultrasound audio. 

\systemname{} builds on this capability, but instead transmits purposefully curated audio to earphone receivers for targeted consumption.

\subsection{Earphone-Based Sound Spheres}
Commercial hearables, such as the \textit{AirPods Pro}, incorporate spatial audio features that simulate immersive, three-dimensional sound environments by dynamically rendering audio relative to the user's head position \cite{apple2025spatialaudio}. Recent research has extended this concept beyond entertainment to enhance situational audio control. For instance, \citet{chen2024hearable} introduced ``sound bubbles'', allowing users to define personal audio zones with radii ranging from 1 to 2 meters. Within these zones, hearables selectively permit audio transmission, effectively isolating the user from external sound sources outside the defined perimeter. In parallel, \citet{veluri2024look} proposed an approach for selective hearing in social settings, where users can focus on a speaker for several seconds to enroll them as a preferred source, enabling the hearable to prioritize their voice while suppressing others.

\systemname{} extends this idea by externalizing spatial audio rendering into the environment. Rather than simulating zones in the earphones, it uses ultrasound to create spatially anchored sound spheres, which are demodulated for position-based access to personalized audio.

\begin{figure*}[!t]
    \centering
    \includegraphics[width=\linewidth,trim=0cm 6.5cm 0cm 6.5cm]{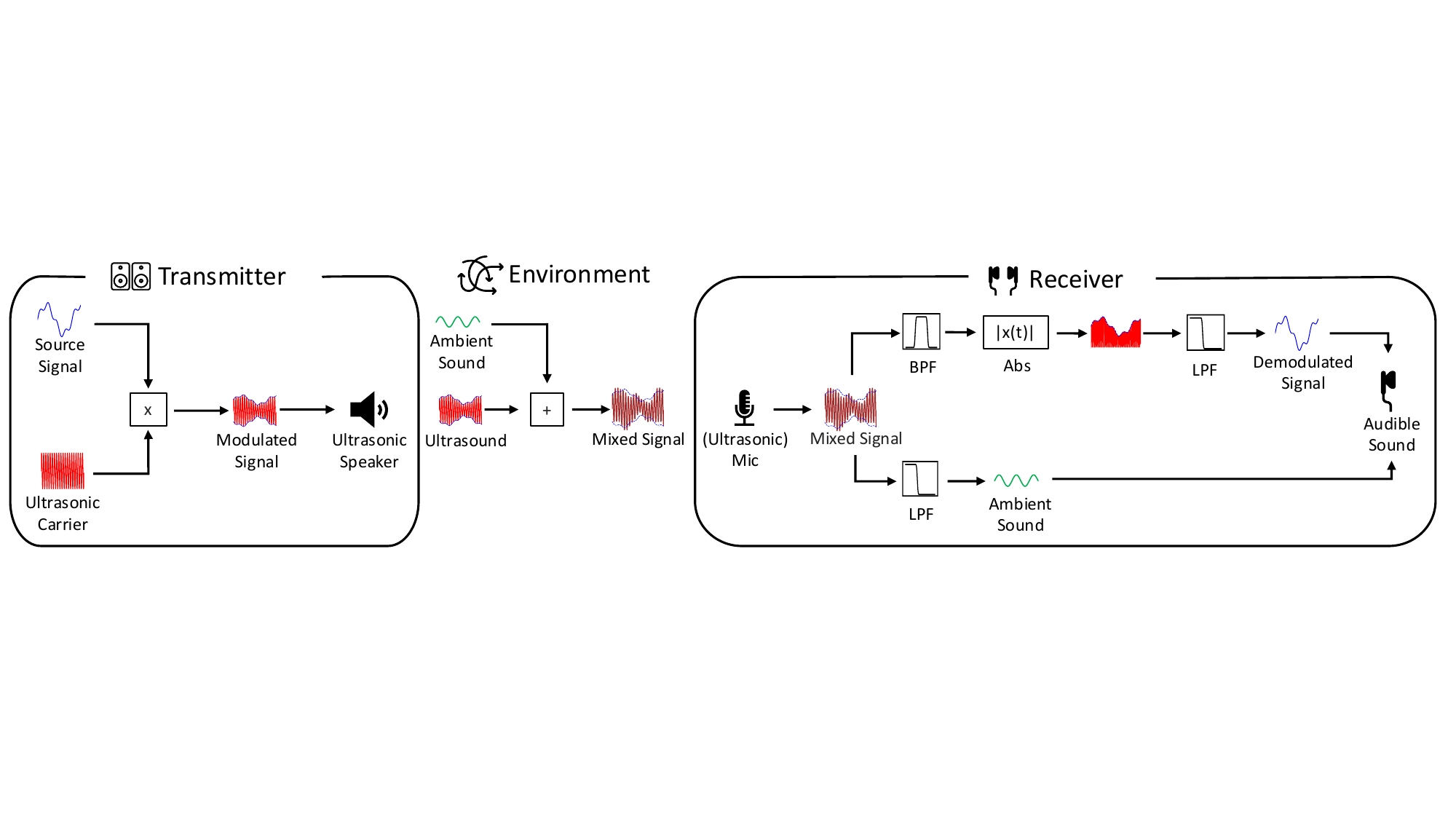}
    \caption{The original signal is modulated and transmitted via a speaker, where it becomes mixed with environmental sound. The receiver captures the combined signal, separates the ultrasonic signal from the noise, and demodulates it. This is achieved by first applying a band-pass filter to isolate the desired ultrasonic channel, followed by a low-pass filter applied to the absolute value of the signal for demodulation into audible sound. Finally, the demodulated signal is reintegrated with the environmental sound to preserve spatial and situational awareness.}
    \label{fig:chain}
\end{figure*}


\section{\systemname{}: Localized, Multi-Channel Sound Spheres via Ultrasound}

\systemname{} creates localized, user-specific audio delivery using ultrasound and wearable earphones. The system operates entirely radio transmission free: audio signals are modulated using amplitude modulation (AM) with an ultrasonic carrier frequency and are continuously emitted by the speaker. Sound becomes perceptible only to users wearing earables and who are standing inside the acoustic beam. We make use of the fact that high-frequency sound waves travel in a more directed manner to achieve a restriction of sound propagation. Since the signal still travels at the speed of sound, the interaural time difference (ITD) is preserved and the user will naturally perceive the location of the sound source.
Using different carrier frequencies, we enable multiple users to receive different content simultaneously in the same physical space, without interference or awareness by others.

The \systemname{} system consists of two main components: a single-speaker ultrasonic transmitter and two earphone receivers. For the transmitter, we use an off-the-shelf, standard high-fidelity bookshelf speaker readily equipped with a tweeter capable of outputting ultrasonic frequencies (~20–55\,kHz). We drive it using a Scarlett Focusrite 2i2 at an output sampling rate of 96\,kHz, which is connected to an standard high-fidelity amplifier. 
No special modifications were made to the speakers; the only requirement is that their high-frequency response, along with the driver chain, is capable of handling the ultrasonic carrier frequencies. In our prototype, we found that many consumer audio tweeters, which often extend into the lower ultrasonic range, can serve this purpose. This means that our system can be realized without investing in specialized ultrasonic emitters. 
For the receiver end, we leverage the open-source OpenEarable 2.0 platform \cite{roddiger2025openearable}, which gives low-level access to the audio processing capabilities of modern noise-canceling earphones. No other special infrastructure is required.

\begin{figure}[!h]
  \centering
  \includegraphics[width=0.9\linewidth,clip,trim=5cm 1cm 5cm 2cm]{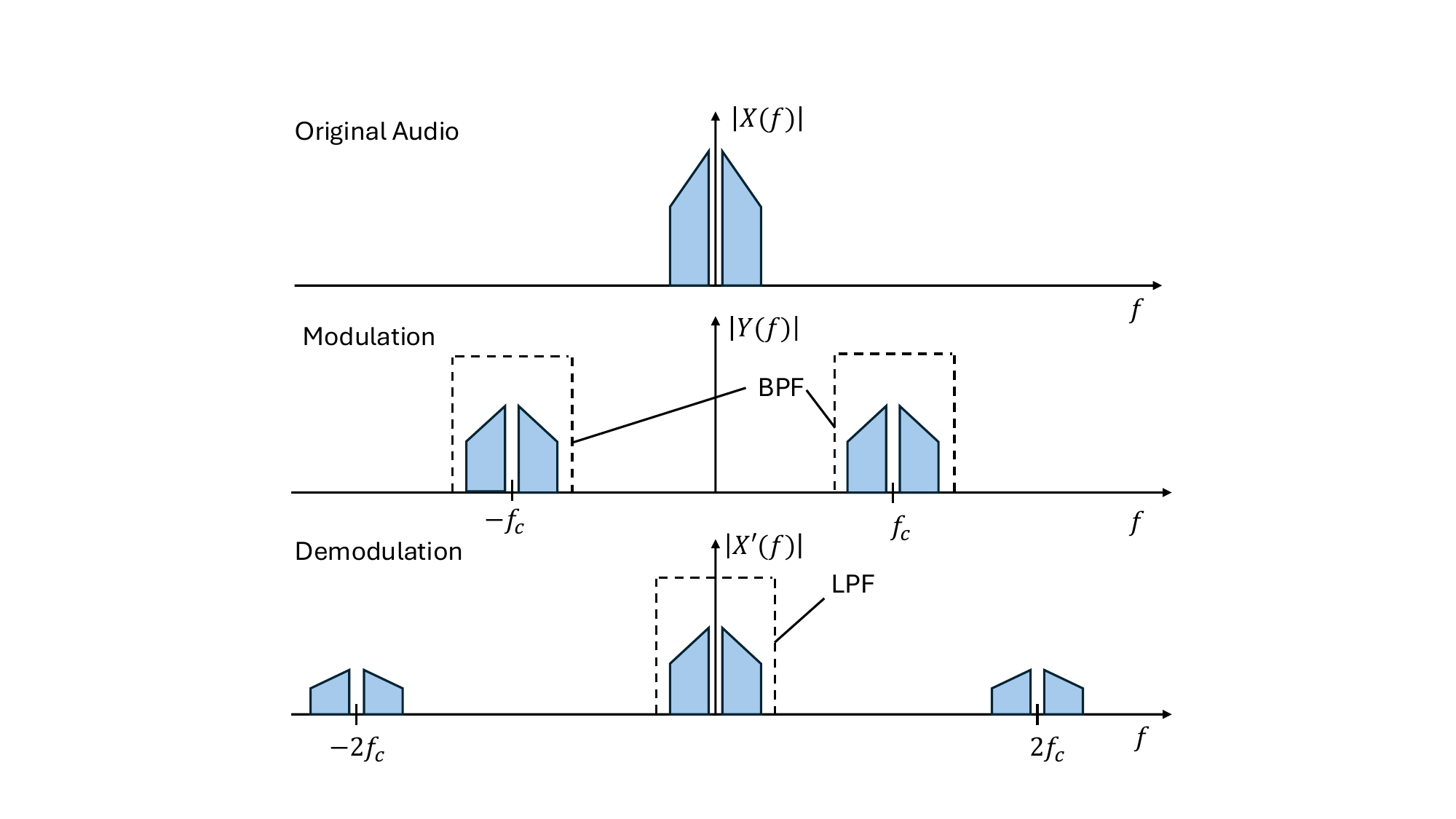}
  \caption{Illustration of the magnitude spectrum in the frequency domain showing modulation and demodulation.}
  \label{fig:spectrum}
\end{figure}

\subsection{Modulation}
To effectively shift the audio signal into the ultrasonic range, we use amplitude modulation (AM). AM is a well-established technique, primarily known from radio signal transmission. However, this method is not limited to electromagnetic waves; it can also be applied to any oscillatory signal, including sound waves.

To restrict the signal's bandwidth, a low-pass filter is applied at first. As human speech ranges up to \freqcut{}, we chose this as the high-cut frequency. The signal is then multiplied with the carrier signal of $f_c$ of \freqcarrier{}.
\begin{equation}
    y(t) = [1 + k_a \cdot x(t)] \cdot \cos(2\pi f_c \cdot t),
\end{equation}
This results in the original signal $X(f)$ being shifted by $f_c$ inside the spectrum, as illustrated in \autoref{fig:spectrum}. We chose a carrier frequency of \freqcarrier{} to ensure that the lower end of the frequency band of the modulated signal $Y(f)$ stays well outside the audible range with 26\,kHz.

\subsection{Demodulation}

To make the modulated ultrasonic signal audible again, we need to demodulate it inside the earphone. To achieve this, the modulated signal is usually multiplied with the carrier signal again and then low-pass filtered afterwards.
\begin{equation}
\begin{split}
    x'(t) =& y(t) \cdot \cos(2\pi f_c \cdot t) \\
    =& \frac{1 + k_a \cdot x(t)}{2} + \frac{1 + k_a \cdot x(t)}{2} \cdot \cos(4 \pi f_c \cdot t) \\
    \hat{x}(t) =& \frac{2 \cdot LPF\{x'(t)\} - 1}{k_a},
\end{split}
\end{equation}
However, this approach requires the receiver to retrieve the phase-synchronized carrier signal. To achieve this, the carrier must typically be recovered from the modulated signal, e.g. using a phase-locked loop (PLL).
As a simple alternative, the absolute value of the modulated signal can instead be used. This removes the need for explicit carrier recovery and will still provide sufficient robustness against noise for our use-case. The audio signal can thus be recovered by the following equation \eqref{eq:demod_abs} using a custom amplification factor $A$.
\begin{equation}
    \tilde{x}(t) = 2A \cdot BPF\{\abs{y(t)}\}
    \label{eq:demod_abs}
\end{equation}

\subsection{Implementation}

On the transmitter side, the audio file is first low-pass filtered and then modulated using MATLAB. To support multiple audio channels, the individual modulated signals are combined into a single composite signal. The result is then exported as a WAV file with a sampling rate of 96\,kHz. This file can be simply played back through the connected speaker system.
For implementation of the receiver, we adjust the routing and filterbanks of the OpenEarable 2.0's DSP (ADAU1860). To allow proper sampling of ultrasonic signals, a sample rate of 192\,kHz was used. The system contains ultrasonic-capable PDM microphones of type Knowles SPH0641LU4H-1 allowing to capture signals of up to 80kHz.

\begin{figure}[t]
    \centering
    \includegraphics[width=\linewidth,page=5,clip,trim=0cm 8.3cm 14cm 0cm]{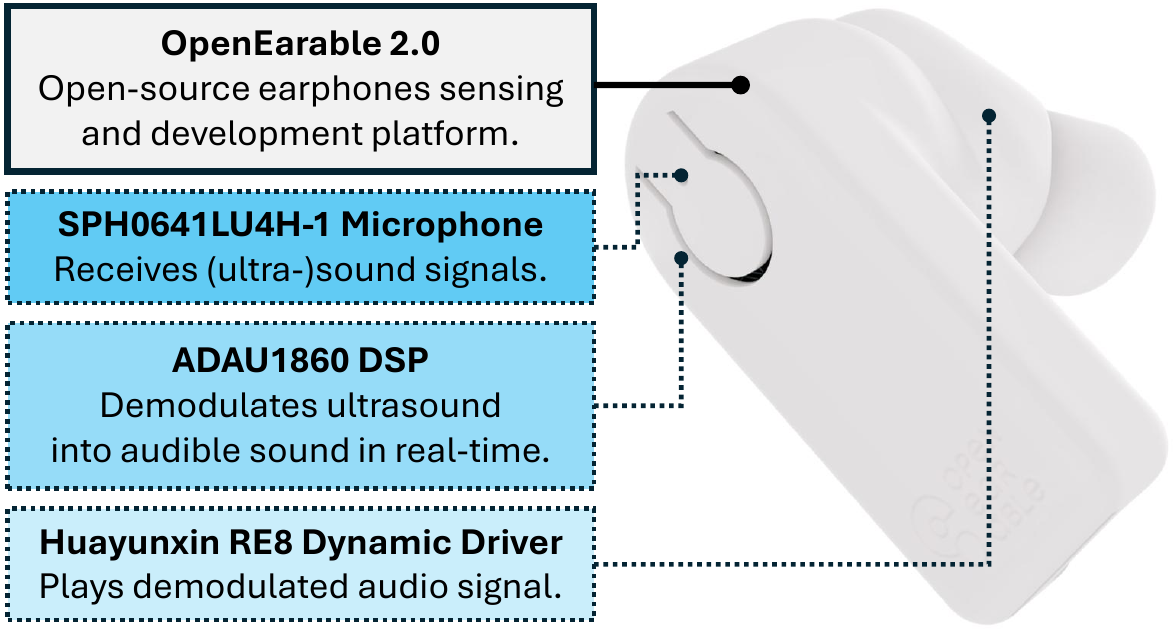}
    \caption{\systemname{} leverages the open-source OpenEarable 2.0 \cite{roddiger2025openearable} and its edge audio processing capabilities, resembling those of standard ANC earphones.}
    \label{fig:enter-label}
\end{figure}

The DSP pipeline comprises three filter stages. The first stage applies a bell-curve equalizer to compensate for imbalances in the ultrasonic frequency response. Next, a Chebyshev Type II band-pass filter isolates the frequency band of interest. This filter type was selected for its steep roll-off and minimal pass-band distortion. As explained above, the absolute value of the signal is then calculated and passed through a low-pass filter to eliminate the high-frequency components. Finally, the signal is amplified and processed through a limiter for dynamic range compression, which helps suppress unwanted clicking sounds from extraneous sources leaking into the ultrasonic band, before being sent to the DAC connected to the dynamic driver. The entire chain is illustrated in \autoref{fig:chain}.

\section{Demonstration}
\begin{figure}[t]
     \centering
    \includegraphics[width=\linewidth]{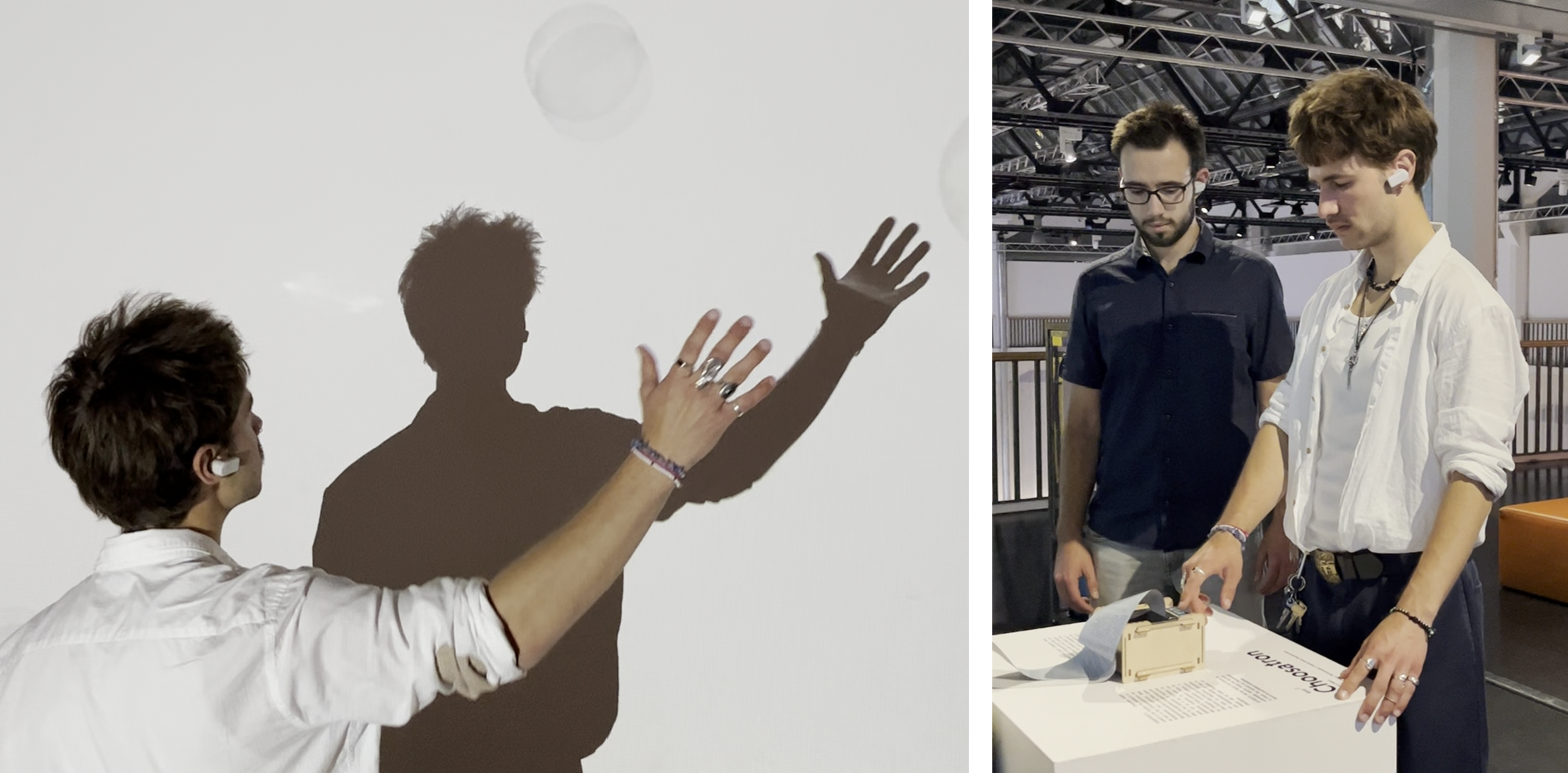}
    \caption{(Left) A visitor experiencing a localized UltrasonicSphere soundscape; (Right) Two visitors looking at an exhibit surrounded by an UltrasonicSphere, each decoding the audio narration to their desired language.}
    \label{fig:exhibit}
\end{figure}
\noindent We demonstrate \systemname{} through an exhibit that resembles a typical scenario at a museum (see \autoref{fig:exhibit}), which lets attendees walk through different localized \systemname{}, each providing multiple audio channels. 
The setup includes two ultrasonic speakers, each positioned at a separate exhibit. 
Both speakers continuously broadcast exhibit-related audio content. One exhibit is accompanied by a soundscape, the other exhibit is supported by related information material on two audio channels: one in English and one in German. 
The spatially located speakers enable simultaneous, location-specific delivery.

Participants of the demo equipped with demodulation earphones of \systemname{} can hear the content by simply walking up to the exhibits. 
They will be able to experience how the ultrasonic signals are turned directly into audible sounds inside the ear canal that can be spatially associated with the location of the exhibit.
Further, the demodulation earphones are fully acoustically transparent. 
This will allow demo participants to remain aware of their surroundings and interact naturally within the exhibition space.
Lastly, participants will be able to switch between the German and English channels by tapping the earphones.
Demo attendees will also be invited to explore the absence of audible sounds when listening to \systemname{} without wearing the required earable receivers.

\section{Conclusion}
We presented \systemname{}, a lightweight system that enables spatially localized and personalized audio delivery using ultrasound and earable devices. By leveraging off-the-shelf speakers and open-source wearable hardware, \systemname{} allows users to hear context-specific content, such as ambient soundscapes or exhibit descriptions, without isolating them from their environment, while operating fully standalone and radio transmission free.

Our demonstration shows that selective audio access can be achieved through a combination of spatial orientation and ultrasonic signal decoding, enabling intuitive and unobtrusive interaction. The system supports user-controlled channel switching (e.g., language selection) via simple ear tap gestures, and is well suited for use in public spaces such as exhibitions or galleries.


\bibliographystyle{ACM-Reference-Format}
\bibliography{references}


\begin{thebibliography}{13}


\ifx \showCODEN    \undefined \def \showCODEN     #1{\unskip}     \fi
\ifx \showISBNx    \undefined \def \showISBNx     #1{\unskip}     \fi
\ifx \showISBNxiii \undefined \def \showISBNxiii  #1{\unskip}     \fi
\ifx \showISSN     \undefined \def \showISSN      #1{\unskip}     \fi
\ifx \showLCCN     \undefined \def \showLCCN      #1{\unskip}     \fi
\ifx \shownote     \undefined \def \shownote      #1{#1}          \fi
\ifx \showarticletitle \undefined \def \showarticletitle #1{#1}   \fi
\ifx \showURL      \undefined \def \showURL       {\relax}        \fi
\providecommand\bibfield[2]{#2}
\providecommand\bibinfo[2]{#2}
\providecommand\natexlab[1]{#1}
\providecommand\showeprint[2][]{arXiv:#2}

\bibitem[{Apple Inc.}(2025)]%
        {apple2025spatialaudio}
\bibfield{author}{\bibinfo{person}{{Apple Inc.}}} \bibinfo{year}{2025}\natexlab{}.
\newblock \bibinfo{title}{Control Spatial Audio and head tracking}.
\newblock
\urldef\tempurl%
\url{https://support.apple.com/en-en/guide/airpods/dev00eb7e0a3/web}
\showURL{%
\tempurl}
\newblock
\shownote{Accessed: 2025-05-24}.


\bibitem[Betlehem et~al\mbox{.}(2015)]%
        {betlehem2015personal}
\bibfield{author}{\bibinfo{person}{Terence Betlehem}, \bibinfo{person}{Wen Zhang}, \bibinfo{person}{Mark~A Poletti}, {and} \bibinfo{person}{Thushara~D Abhayapala}.} \bibinfo{year}{2015}\natexlab{}.
\newblock \showarticletitle{Personal sound zones: Delivering interface-free audio to multiple listeners}.
\newblock \bibinfo{journal}{\emph{IEEE Signal Processing Magazine}} \bibinfo{volume}{32}, \bibinfo{number}{2} (\bibinfo{year}{2015}), \bibinfo{pages}{81--91}.
\newblock


\bibitem[Chen et~al\mbox{.}(2024)]%
        {chen2024hearable}
\bibfield{author}{\bibinfo{person}{Tuochao Chen}, \bibinfo{person}{Malek Itani}, \bibinfo{person}{Sefik~Emre Eskimez}, \bibinfo{person}{Takuya Yoshioka}, {and} \bibinfo{person}{Shyamnath Gollakota}.} \bibinfo{year}{2024}\natexlab{}.
\newblock \showarticletitle{Hearable devices with sound bubbles}.
\newblock \bibinfo{journal}{\emph{Nature Electronics}} (\bibinfo{year}{2024}), \bibinfo{pages}{1--12}.
\newblock


\bibitem[Fan et~al\mbox{.}(2023)]%
        {fan2023apg}
\bibfield{author}{\bibinfo{person}{Xiaoran Fan}, \bibinfo{person}{David Pearl}, \bibinfo{person}{Richard Howard}, \bibinfo{person}{Longfei Shangguan}, {and} \bibinfo{person}{Trausti Thormundsson}.} \bibinfo{year}{2023}\natexlab{}.
\newblock \showarticletitle{Apg: Audioplethysmography for cardiac monitoring in hearables}. In \bibinfo{booktitle}{\emph{Proceedings of the 29th Annual International Conference on Mobile Computing and Networking}}. \bibinfo{pages}{1--15}.
\newblock


\bibitem[Gao et~al\mbox{.}(2019)]%
        {gao2019earecho}
\bibfield{author}{\bibinfo{person}{Yang Gao}, \bibinfo{person}{Wei Wang}, \bibinfo{person}{Vir~V Phoha}, \bibinfo{person}{Wei Sun}, {and} \bibinfo{person}{Zhanpeng Jin}.} \bibinfo{year}{2019}\natexlab{}.
\newblock \showarticletitle{EarEcho: Using ear canal echo for wearable authentication}.
\newblock \bibinfo{journal}{\emph{Proceedings of the ACM on Interactive, Mobile, Wearable and Ubiquitous Technologies}} \bibinfo{volume}{3}, \bibinfo{number}{3} (\bibinfo{year}{2019}), \bibinfo{pages}{1--24}.
\newblock


\bibitem[{Holosonics}(2025)]%
        {Holosonics}
\bibfield{author}{\bibinfo{person}{{Holosonics}}.} \bibinfo{year}{2025}\natexlab{}.
\newblock \bibinfo{title}{{Audio Spotlight by Holosonics}}.
\newblock \bibinfo{howpublished}{\url{https://www.holosonics.com/}}.
\newblock
\newblock
\shownote{Accessed: 2025-05-22}.


\bibitem[Jin et~al\mbox{.}(2021)]%
        {jin2021sonicasl}
\bibfield{author}{\bibinfo{person}{Yincheng Jin}, \bibinfo{person}{Yang Gao}, \bibinfo{person}{Yanjun Zhu}, \bibinfo{person}{Wei Wang}, \bibinfo{person}{Jiyang Li}, \bibinfo{person}{Seokmin Choi}, \bibinfo{person}{Zhangyu Li}, \bibinfo{person}{Jagmohan Chauhan}, \bibinfo{person}{Anind~K Dey}, {and} \bibinfo{person}{Zhanpeng Jin}.} \bibinfo{year}{2021}\natexlab{}.
\newblock \showarticletitle{SonicASL: An acoustic-based sign language gesture recognizer using earphones}.
\newblock \bibinfo{journal}{\emph{Proceedings of the ACM on Interactive, Mobile, Wearable and Ubiquitous Technologies}} \bibinfo{volume}{5}, \bibinfo{number}{2} (\bibinfo{year}{2021}), \bibinfo{pages}{1--30}.
\newblock


\bibitem[Mei et~al\mbox{.}(2022)]%
        {mei2022parametric}
\bibfield{author}{\bibinfo{person}{Shangming Mei}, \bibinfo{person}{Hui Xu}, \bibinfo{person}{Yihua Hu}, \bibinfo{person}{Mohammed Alkahtani}, {and} \bibinfo{person}{Yangang Wang}.} \bibinfo{year}{2022}\natexlab{}.
\newblock \showarticletitle{The Parametric Array Speaker: A Review}. In \bibinfo{booktitle}{\emph{International Joint Conference on Energy, Electrical and Power Engineering}}. Springer, \bibinfo{pages}{1254--1271}.
\newblock


\bibitem[Ochiai et~al\mbox{.}(2017)]%
        {ochiai2017holographic}
\bibfield{author}{\bibinfo{person}{Yoichi Ochiai}, \bibinfo{person}{Takayuki Hoshi}, {and} \bibinfo{person}{Ippei Suzuki}.} \bibinfo{year}{2017}\natexlab{}.
\newblock \showarticletitle{Holographic whisper: Rendering audible sound spots in three-dimensional space by focusing ultrasonic waves}. In \bibinfo{booktitle}{\emph{proceedings of the 2017 CHI conference on human factors in computing systems}}. \bibinfo{pages}{4314--4325}.
\newblock


\bibitem[R{\"o}ddiger et~al\mbox{.}(2025)]%
        {roddiger2025openearable}
\bibfield{author}{\bibinfo{person}{Tobias R{\"o}ddiger}, \bibinfo{person}{Michael K{\"u}ttner}, \bibinfo{person}{Philipp Lepold}, \bibinfo{person}{Tobias King}, \bibinfo{person}{Dennis Moschina}, \bibinfo{person}{Oliver Bagge}, \bibinfo{person}{Joseph~A Paradiso}, \bibinfo{person}{Christopher Clarke}, {and} \bibinfo{person}{Michael Beigl}.} \bibinfo{year}{2025}\natexlab{}.
\newblock \showarticletitle{OpenEarable 2.0: Open-Source Earphone Platform for Physiological Ear Sensing}.
\newblock \bibinfo{journal}{\emph{Proceedings of the ACM on Interactive, Mobile, Wearable and Ubiquitous Technologies}} \bibinfo{volume}{9}, \bibinfo{number}{1} (\bibinfo{year}{2025}), \bibinfo{pages}{1--33}.
\newblock


\bibitem[Veluri et~al\mbox{.}(2024)]%
        {veluri2024look}
\bibfield{author}{\bibinfo{person}{Bandhav Veluri}, \bibinfo{person}{Malek Itani}, \bibinfo{person}{Tuochao Chen}, \bibinfo{person}{Takuya Yoshioka}, {and} \bibinfo{person}{Shyamnath Gollakota}.} \bibinfo{year}{2024}\natexlab{}.
\newblock \showarticletitle{Look once to hear: Target speech hearing with noisy examples}. In \bibinfo{booktitle}{\emph{Proceedings of the 2024 CHI Conference on Human Factors in Computing Systems}}. \bibinfo{pages}{1--16}.
\newblock


\bibitem[Watanabe and Terada(2023)]%
        {watanabe2023ultrasonicwhisper}
\bibfield{author}{\bibinfo{person}{Hiroki Watanabe} {and} \bibinfo{person}{Tsutomu Terada}.} \bibinfo{year}{2023}\natexlab{}.
\newblock \showarticletitle{UltrasonicWhisper: Ultrasound Can Generate Audible Sound in Your Hearable}. In \bibinfo{booktitle}{\emph{Proceedings of the 2023 ACM International Symposium on Wearable Computers}}. \bibinfo{pages}{132--134}.
\newblock


\bibitem[Zhong et~al\mbox{.}(2025)]%
        {zhong2025audible}
\bibfield{author}{\bibinfo{person}{Jia-Xin Zhong}, \bibinfo{person}{Jun Ji}, \bibinfo{person}{Xiaoxing Xia}, \bibinfo{person}{Hyeonu Heo}, {and} \bibinfo{person}{Yun Jing}.} \bibinfo{year}{2025}\natexlab{}.
\newblock \showarticletitle{Audible enclaves crafted by nonlinear self-bending ultrasonic beams}.
\newblock \bibinfo{journal}{\emph{Proceedings of the National Academy of Sciences}} \bibinfo{volume}{122}, \bibinfo{number}{12} (\bibinfo{year}{2025}), \bibinfo{pages}{e2408975122}.
\newblock


\end{thebibliography}


\end{document}